\documentclass[epj]{svjour}
\usepackage{graphics}
\begin{document}
\title{Today's View on Strangeness}
\author{John Ellis}
%
\offprints{}          
\institute{CERN, CH-1211 Geneva 23, Switzerland}
\date{Received: date / Revised version: date}
%
\abstract{
There are several different experimental indications, such as the
pion-nucleon $\Sigma$ term and polarized deep-inelastic scattering, which
suggest that the nucleon wave function contains a hidden $s {\bar s}$
component. This is expected in chiral soliton models, which also predicted
the existence of new exotic baryons that may recently have been observed.
Another hint of hidden strangeness in the nucleon is provided by copious
$\phi$ production in various $N {\bar N}$ annihilation channels, which may
be due to evasions of the Okubo-Zweig-Iizuka rule. One way to probe the
possible polarization of hidden $s {\bar s}$ pairs in the nucleon may be
via $\Lambda$ polarization in deep-inelastic scattering.
} 
\maketitle
\begin{center}
CERN-PH-TH/2004-231 ~~~~~~~~~~~~ hep-ph/0411369
\end{center}

\section{How Strange is the Nucleon?}
\label{sec:1}
Some people might argue that this is a ``strange" question: why should the
nucleon be strange at all - after all, is
it not just made out of three up and down quarks? We should not jump to
such a na\"ive conclusion. For a start,
even the vacuum is strange: chiral symmetry for $\pi, K$ mesons tells us 
that~\cite{svac}
$$
<0 | \bar{s} s|0> = (0.8 \pm 0.1) <0 | \bar{q} q|0>.
$$
This hidden strangeness cannot be expected to disappear when one inserts a
set of three quark `coloured test
charges' into the vacuum. Moreover, hidden strangeness will be generated in
perturbative QCD:
$$
{\rm quark} \, \to {\rm gluon} \, \to {\bar s} s  \,{\rm pair}.
$$
There are also non-perturbative mechanisms for generating ${\bar s} s$ pairs
in the nucleon, such as instanton effects~\cite{GtH}.

Another objection to this `strange' question is the fact that (at least
some) experiments do not see very much
strangeness in the nucleon. For example, CCFR measures a strange momentum
fraction: $P_s = 4 \%$ at $Q^2 = 20
{\rm GeV}^2$~\cite{CCFR}, the HAPPEX measurement of a combination of 
strange electric
and magnetic form factors gives a small
value: $G_E + 0.39 G_M = 0.025 \pm 0.020 \pm 0.014$ at $Q^2 = 0.48 {\rm
GeV}^2$~\cite{HAPPEX}, SAMPLE finds a small
strange contribution to the nucleon magnetic moment: $-0.1 \pm 5.1 
\%$~\cite{SAMPLE}, and the A4 Collaboration finds small strange contribution
to another combination of form factors: $G_E + 0.225 G_M = 0.039 \pm 
0.034$~\cite{A4}.

On the other hand, a few experiments indicate quite large matrix elements
for some hidden-strangeness operators. One
prominent example is the $\pi$-nucleon $\Sigma$ term, whose value is
related to the strange scalar density:
$$
y = \frac{2 < p|\bar{s}s |p>} {<p|\bar{u} u |p> + <p | \bar{d}d | p>} .
$$
Two recent determinations of the $\pi$-nucleon $\Sigma$ term have
found large values~\cite{sigma}
:
$$
\Sigma = 64 \pm 8, \; (79 \pm 7) \, {\rm MeV}
$$
corresponding to large values of $y = 1 - \sigma_0/\Sigma$, where octet 
baryon mass differences give $\sigma_0 = 36 \pm 7$ MeV~\cite{Knecht} and 
hence $y \sim 0.5$. Another example is the strange spin of the
nucleon: a na\"{i}ve interpretation of
measurements of polarized deep-inelastic structure functions would 
yield~\cite{spol}:
$$
\Delta s = dx[s_\uparrow (x) - s_\downarrow (x) + \bar{s}_\uparrow(x) -
\bar{s}_\downarrow(x)] = -0.10 \pm 0.02.
$$
On the other hand, HERMES measurements of single-particle inclusive
particle production have been interpreted as
indicating that~\cite{sinc}
$$
\Delta s = dx[s_\uparrow (x) - s_\downarrow (x) + \bar{s}_\uparrow(x) -
\bar{s}_\downarrow(x)] = 0.03 \pm 0.03 \pm
0.01.
$$
However, this estimate has been questioned on the grounds that corrections
to independent fragmentation may be
large \cite{Aram}. The
overall picture is that
hidden-strangeness matrix elements in the nucleon may be small or large,
depending on the $J^{PC}$ quantum numbers
carried by the $\bar{s} s$ pair, which is quite compatible with theoretical
ideas~\cite{KL}.

\begin{figure}[htb]
\vspace{0.5cm}
\resizebox{0.45\textwidth}{!}{%
\includegraphics{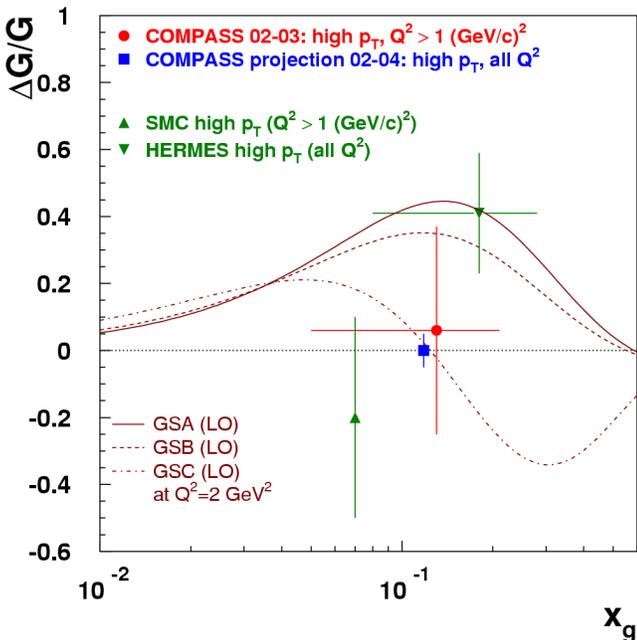}
}
\caption{Comparison of recent determinations of $\Delta G$ by 
HERMES~\protect\cite{HERMES}, SMC~\protect\cite{SMC} 
and COMPASS~\protect\cite{COMPASS}.
}
\label{fig:DeltaG}
\end{figure}

Even if one accepts the first estimate of $\Delta s$, there has been an
argument about its interpretation, based on
the observation that in one regularization scheme  $\Delta s$ gets a large
contribution from gluons $\Delta g$:
$\tilde{\Delta} s = \Delta s  - (\alpha_s / \pi)\Delta g$: perhaps the
`bare' $\Delta s$  vanishes, and $\Delta g$ is
large and positive~\cite{AR}? Since the
$\Delta g$ correction is scheme-dependent, one may wonder how well defined
it is~\cite{Jaffe}. However, this suggestion has at least raised
the profile of the interesting question how large $\Delta g$ may be. 
A first measurement of the gluon
polarization was reported by FNAL experiment E581/704~\cite{FNAL}, 
measuring $\pi^0$ production at
high $p_T$, and Fig.~\ref{fig:DeltaG} shows recent measurements of 
$\Delta G$. HERMES find~\cite{HERMES}
$$
<\Delta G / G> = 0.41 \pm 0.18 \, {\rm (stat)} \pm 0.03 \, {\rm (syst)}
$$
for $0.06 < x_G < 0.28$, and the SMC finds~\cite{SMC}
$$
\Delta G / G = - 0.20 \pm 0.28 \pm 0.10
$$
for an average longitudinal momentum fraction $<\eta \equiv x_G (1 + 
{\hat s}/Q^2)> = 0.07$
using the asymmetry in hadron-pair production at high $p_T$. 
Most recently,
COMPASS has announced a new determination, again via the
asymmetry in hadron-pair production at high $p_T$~\cite{COMPASS}:
\begin{equation}
\Delta G / G = 0.06 \pm 0.31 \pm 0.06 
\end{equation}
at an average $<x_G> = 0.13$,
and PHENIX is preparing a new determination via the double
helicity asymmetry in $pp \to \pi^0$ at high
$p_T$. However, all
these measurements have large
uncertainties, both systematic and statistical. For the time being, 
there is no strong indication that $\Delta G$ is large, and even its
sign must still be regarded as an open question.

\section{Models of the Nucleon}
\label{sec:2}

In the last millennium, the na\"{i}ve quark model (NQM)~\cite{Kokk} held 
pride 
of place. It
envisages the nucleon as composed of three
constituent quarks $Q$, each with mass $M_Q \sim 300$ MeV, in a
non-relativistic wave function. Any additional $q
\bar{q}$ pairs are thought to be generated perturbatively, and few of them
are expected to be $s \bar{s}$ pairs.
Baryons sit in the usual non-exotic SU(3) multiplets, and the combination
of a UUD or UDD wave function with meagre
pair creation explains the Okubo-Zweig-Iizuka (OZI) rule~\cite{OZI}. The 
proton spin
is simply the algebraic sum of valence
constituent-quark spins, which add up to $1/2$.

Chiral soliton models~\cite{Skyrme} provide an alternative viewpoint for 
the new
millennium. They are based on the observation
that the intrinsic masses of the (current) quarks defined at short
distances are much smaller: $m_{u,d} \sim$ few
MeV, $m_s \sim 100$ MeV. Hence the quarks should be treated
relativistically, and there are many intrinsic $q
\bar{q}$
pairs in the nucleon wave function, which are treated as clouds of meson
fields. In this picture, low-lying exotic
SU(3) multiplets are predicted~\cite{exotics}, as are evasions of the OZI 
rule due to the
copious $s \bar{s}$ pairs in the nucleon.
Moreover, the nucleon spin is obtained from orbital angular momentum, the
sum of the quark spins vanishes in the
limit of vanishing quark masses and a large number of colours, and the $s
\bar{s}$ pairs are polarized~\cite{BEK}.

In the chiral soliton model, baryons are constructed as clouds of $\pi, K$,
and $\eta_8$ mesons, and the presence of
the latter is one way to understand the copious $s \bar{s}$ pairs in the
baryon wave function. Exotic baryons are
expected as excitations of the meson cloud with non-trivial SU(3)
transformation properties, which can also be
interpreted as excitations of the $q \bar{q}$ sea in the baryon. The baryon
spin is due to the coherent rotation of
this meson cloud, motivating the interpretation of the baryon spin as
orbital angular momentum, and requiring the
$s \bar{s}$ pairs to be polarized.

Specifically, in the limit of massless quarks and a large number of
colours, the meson cloud contains no SU(3)
flavour-singlet $\eta_0$ mesons, and nor do the $\pi, K$, and $\eta_8$
mesons present have any coupling to the
$\eta_0$. Since axial-current matrix elements are related in the chiral
limit to pseudoscalar-meson couplings, the
absence of the
$\eta_0$ implies that the SU(3)-singlet axial-current matrix element
between baryons also vanishes. Classically, this
matrix element is in turn related to the sum of the quark spins in the
baryon, which therefore vanishes. Since the
sum of the $u$ and $d$ quark contributions to the proton spin is positive
and does not vanish, there must be a
negative, non-zero strange contribution that cancels them~\cite{BEK}.

The presence of a non-trivial $q \bar{q}$ sea in the nucleon suggests that
there may exist baryons with `exotic'
quantum numbers that cannot be explained in terms of na\"{i}ve three-quark
wave functions. It is surely too
na\"{i}ve to imagine that, if one places three quarks in a vacuum
containing many $q \bar{q}$ pairs, there will never
be any rearrangement of the $q \bar{q}$ quantum numbers. If the $q \bar{q}$
quantum numbers do not cancel each other
out exactly, the resulting baryonic state will have `exotic' quantum
numbers. In the chiral-soliton language, these
can be thought of as excitations of the meson cloud.

This line of argument led theorists working on chiral solitons to predict
the existence of a relatively light antidecuplet of exotic
baryons~\cite{exotics}, resembling `pentaquark' states in the 
NQM, of which the lowest-lying member would have $ud ud
\bar{s}$ quantum numbers and weigh about 1530 MeV~\cite{DPP}. On the other
hand, the Particle Data Group quoted in 1987 ``\dots the prejudice
against baryons not made of three quarks \dots ", and ceased to consider
their existence.

\begin{figure}
\resizebox{0.45\textwidth}{!}{%
\includegraphics{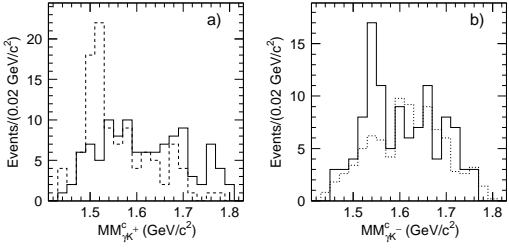}
}
\caption{The left panel shows that the $\Lambda (1520)$ signal can be 
isolated with suitable cuts. The signal for the exotic baryon 
$\Theta^+$ is the solid histogram in 
the right panel, and the dashed histogram is a control 
sample~\protect\cite{Theta1}. 
}
\label{fig:LEPS}
\end{figure}

\begin{figure}
\resizebox{0.45\textwidth}{!}{%
\includegraphics{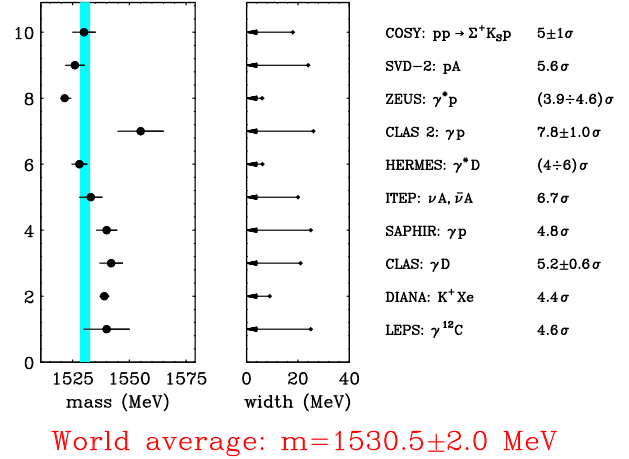}
}
\caption{Compilation of measurements of the $\Theta$ mass and decay 
width~\protect\cite{Karliner}.
}
\label{fig:Close}
\end{figure}

This changed with the report by the LEPS 
Collaboration at Spring-8~\cite{Theta1} of a candidate exotic 
baryon with $udud \bar{s}$ quantum numbers and weighing about 1540 MeV, 
shown in 
Fig.~\ref{fig:LEPS}. This was soon followed by an
avalanche of corroborating evidence from other experiments~\cite{others},
which stimulated considerable theoretical enthusiasm. However, these
observations are somewhat problematic. The masses vary outside the quoted
statistical and systematic errors, as seen in Fig.~\ref{fig:Close}, with 
peaks in $nK^+$ final states
tending to be heavier than those in $pK^0$ final states~\cite{Close}.
Moreover, KN partial-wave analyses require the decay width to be $< 1$
MeV~\cite{Cahn}: this is surprisingly narrow, and some experiments have
reported widths close to their experimental mass resolutions, as also 
seen in Fig.~\ref{fig:Close}.

\begin{figure}[htb]
\resizebox{0.45\textwidth}{!}{%
\includegraphics{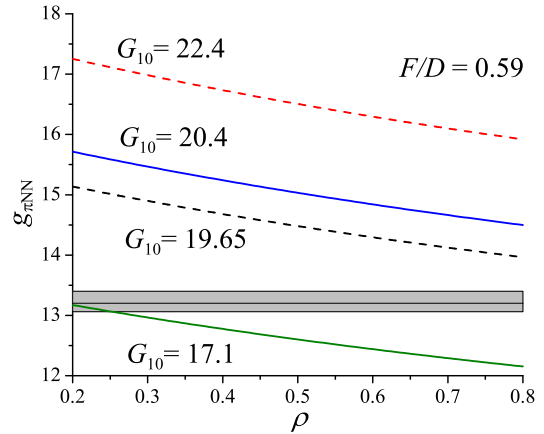}
}
\caption{Chiral-soliton calculations of the $\pi$-nucleon 
coupling depend on the model parameters $G_{10}$ and $\rho$, but may be 
consistent with experiment (shaded)~\protect\cite{EKP}.
}
\label{fig:gpiNN3}
\end{figure}

\begin{figure}[t]
\vspace{1cm}
\resizebox{0.45\textwidth}{!}{%
\includegraphics{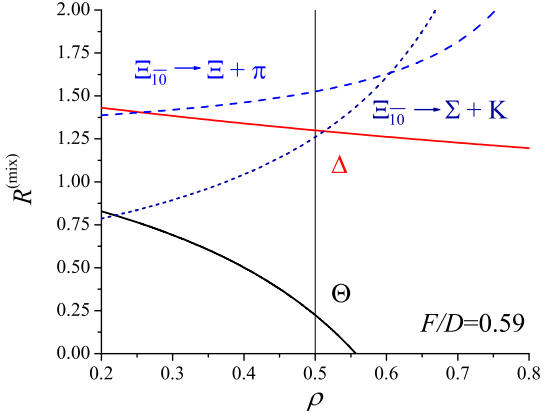}
}
\caption{Corrections to the $\Theta$ coupling tend to reduce its 
decay width, whereas the decays of other antidecuplet baryons are 
not strongly suppressed~\protect\cite{EKP}.
}
\label{fig:rmix2}
\end{figure}

\begin{figure}[b]
\resizebox{0.45\textwidth}{!}{%
\includegraphics{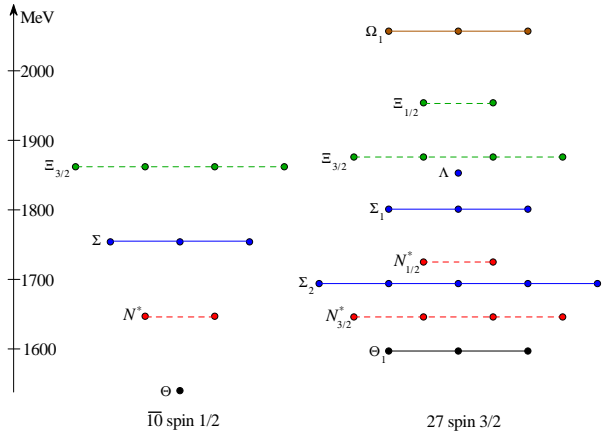}
}
\caption{Spectroscopy of the lowest-lying exotic baryons 
predicted in chiral-soliton models~\protect\cite{EKP}.
}
\label{fig:exotic}
\end{figure}

Nevertheless, the striking evidence in favour of the early chiral-soliton
predictions motivated revisiting them \cite{EKP}. It was soon realized
that the accuracy of the mass prediction~\cite{DPP} was somewhat 
fortuitous, as it
was based on a debatable assignment of another member of the baryon
antidecuplet, and the mass splittings within this multiplet were
calculated using an outdated value for the $\pi$-nucleon $\Sigma$ term.
Using plausible ranges for the chiral soliton moments of inertia that
control the mean excitation energy of antidecuplet baryons, and the more
modern value of the $\Sigma$ term discussed above, there is an uncertainty
in the $\theta^+$ baryon mass of at least 100 MeV. As for the decay width,
although it vanishes in the limit of a large number of colours~\cite{P0},
leading-order calculations with plausible baryon couplings had difficulty
in pushing the decay width below about 10 MeV. We made a detailed study of
SU(3)-symmetry breaking effects on baryon-meson couplings in chiral
soliton models, finding values of the $\pi$-Nucleon and $\pi - \Delta$
couplings that are consistent with experiment, as seen in 
Fig.~\ref{fig:gpiNN3}. These effects tend to
reduce further the $\theta^+$ decay rate, as seen in 
Fig.~\ref{fig:rmix2}, though a width $< 1$ MeV still
seems unlikely~\cite{EKP}. One of the key predictions of chiral soliton
models is the existence of other, more `exotic' baryon multiplets, such as
a 27 and a 35 of SU(3) that are slightly heavier than the antidecuplet. In
particular, there should be a $\theta^{++}$ state weighing $< 100$ MeV
more than the $\theta^+$, as seen in Fig.~\ref{fig:exotic}. It is
difficult to understand how such a
state could have escaped observation in many experiments, but CLAS
data may hint at the existence of such a
state~\cite{CLAS2}.

What would be the implications of `exotic' baryons for our understanding of
strangeness in the nucleon? As has
already been mentioned, the exotic baryon spectrum is sensitive to the
value of the $\pi$-N $\Sigma$ term. Using the
chiral soliton mass formula
\begin{eqnarray}
\frac{m_s}{m} \, \Sigma &  = &
\underbrace{3(4M_{\Sigma} - 3M_{\Lambda} - M_N)} +
\underbrace{4(M_{\Omega} - M_{\Delta})} \nonumber \\
& & \qquad \qquad octet \qquad \qquad \qquad  decuplet \nonumber \\
& - & \underbrace{-4(M_{\Xi 3/2} - M_{\Theta^+})} \nonumber \\
 & &  \quad \quad antidecuplet  \nonumber
\end{eqnarray}
~ \\
~ \\
~ \\
the observation of the $\Xi^{--}$ baryon reported by NA49~\cite{NA49}, if 
confirmed, would correspond to
$$
y = 2 \cdot \frac{<N|s \bar{s} | N>}
{<N|u\bar{u} + d\bar{d}| N>} \approx 0.6 .
$$
This is quite consistent with the direct measurements of the $\pi$-N 
$\Sigma$ term discussed earlier~\cite{sigma}, perhaps lending
some credence to the whole chiral soliton scheme.

\section{OZI Violation or Evasion?}
\label{sec:3}

The Okubo-Zweig-Iizuka (OZI) rule is based on the idea that processes with
disconnected quark lines are suppressed.
As a corollary, it is not possible to produce $s \bar{s}$ mesons in the
interactions of non-strange particles.
Hence, $\phi$ meson production should be due only to the admixture of light
quarks in the $\phi$ wave function,
which is small, since the $\phi$ and $\omega$ mesons are almost ideally
mixed. Generically, one would expect a
production ratio
$$
	R(\phi / \omega) = \tan^2 (\theta - \theta_I) = 4.2 \times 10^{-3}.
$$
 This is not very different from the weighted averages of experimental data
from
$\pi N$ collisions:
$$
	R(\phi / \omega) = (3.30 \pm 0.34) \times 10^{-3} .
$$
The corresponding ratios in NN collisions
$$
	R(\phi / \omega) = (12.78 \pm 0.34) \times 10^{-3}
$$
 and NN collisions
$$
	R(\phi / \omega ) = (14.55  \pm 1.92) \times 10^{-3}
$$
are somewhat larger, but not dramatically big.

\begin{figure}
\resizebox{0.45\textwidth}{!}{%
\includegraphics{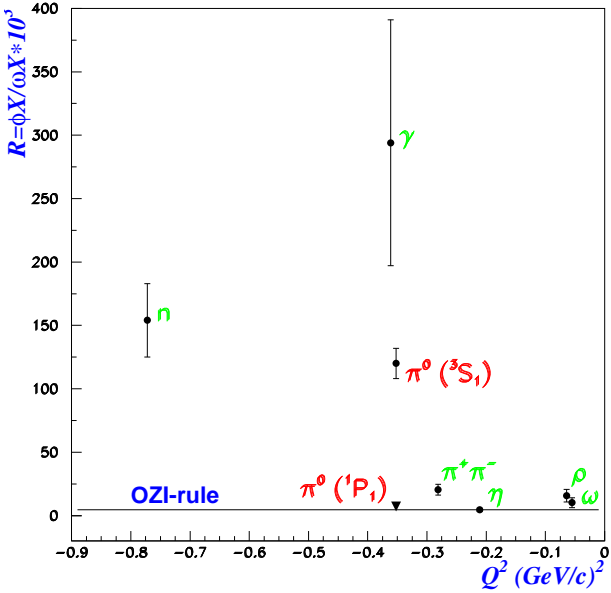}
}
\caption{The ratios of $\phi$ and $\omega$ production in 
association with various other particles
in $N \bar N$ annihilation, as measured at LEAR~\protect\cite{LEAR}, 
often exceed predictions based on the na\"ive OZI rule.
}
\label{fig:ozi}
\end{figure}

\begin{figure}
\resizebox{0.45\textwidth}{!}{%
\includegraphics{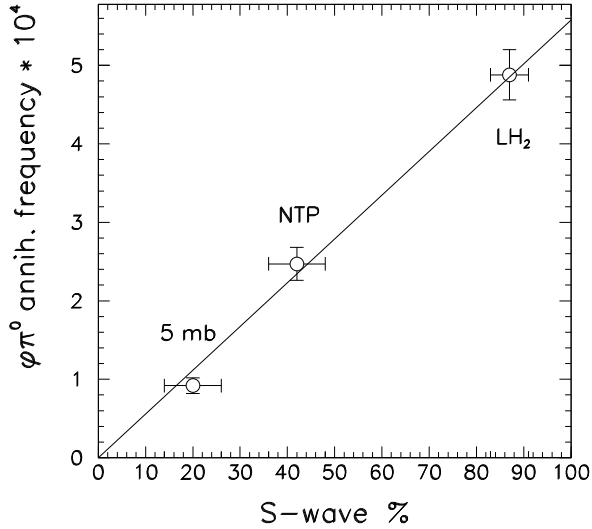}
}
\caption{The annihilation $p \bar p \to \phi \pi^0$ proceeds predominantly 
via the $s$ wave~\protect\cite{LEAR}. 
}
\label{fig:swave}
\end{figure}

\begin{figure}
\resizebox{0.45\textwidth}{!}{%
\includegraphics{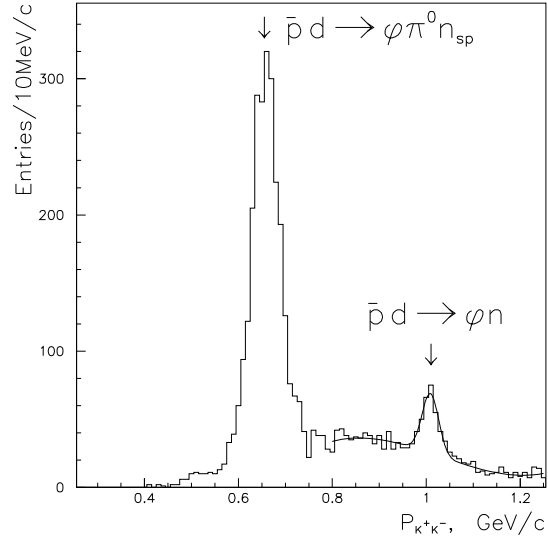}
}
\caption{Signal for the Pontecorvo reaction $\bar p d \to n 
\phi$~\protect\cite{LEAR}.
}
\label{fig:ponte}
\end{figure}

On the other hand, there are large deviations from the na\"{i}ve OZI
relation in data from LEAR experiments on $p \bar{p}$ annihilations,
particularly in the following reactions: $p \bar{p} \to \gamma \phi$, $p
\bar{p} \to \pi \phi$ from the $^3 S_1$ state, and $\bar{p} d \to \phi n$,
as seen in Fig.~\ref{fig:ozi}~\cite{LEAR}. Moreover, the $\phi / \omega$ 
ratio depends
strongly on the initial-state spins of the nucleons and antinucleons, on
their orbital angular momenta, on the momentum transfer and on the
isospin. For example, the partial-wave dependence of annihilations into
$\phi \pi$ is shown in Fig.~\ref{fig:swave}, where we see that $s$-wave 
annihilations
dominate. Another example of a large $\phi / \omega$ ratio is in the
Pontecorvo reaction $\bar{p} d \to \phi n$ shown in Fig.~\ref{fig:ponte}, 
where it is compared with the annihilation process $\bar{p} d \to \pi^0 n$.

The OZI rule could be evaded if there are $s \bar{s}$ pairs in the nucleon
wave function, since new classes of connected quark diagrams could be
drawn for the production of the $\phi$ and other $s \bar{s}$ mesons.
Motivated by the data on polarized deep-inelastic scattering~\cite{EGK},
we have formulated a polarized intrinsic strange- ness
model~\cite{EKKS1,EKKS2}, in which the $s \bar{s}$ pairs in the nucleon
are assumed to have negative polarization, and to be in a relative
$0^{++}$ state, not a $1^{--}$ state as in the na\"{i}ve $\phi$ wave
function. The production of strangeonium states may occur via
rearrangement of the $s$ and $\bar{s}$ in different nucleons, and not via
shake-out from an individual nucleon. Thus, {\it both} nucleons
participate in the production mechanism, and their relative polarization
and orbital angular momentum states are important. In particular, one
would expect the $\phi$ and the $f'_2(1525)$ mesons to be produced more
copiously from spin-triplet initial states than from spin-singlet initial
states, the $\phi$ meson to be produced preferentially from $L = 0$
states, and the $f'_2(1525)$ to be produced preferentially from $L = 1$
states.

This model has led to several correct predictions~\cite{LEAR}. In
nucleon-antinucleon annihilations, the $p \bar{p} \to \pi^0 \phi$ rates
from the $^3S_1$ and $^1P_1$ initial states are in a ratio $\sim 15:1$, in
agreement with the prediction of $L = 0$ dominance. On the other hand, the
$p \bar{p} \to f_2' \pi^0$ rates are in a ratio $ \sim 1:10$, in agreement
with the prediction of $L = 1$ dominance. Moreover, there is evidence that
the mechanisms for $\phi$ and $\omega$ production are different: the
$^1P_1$ fractions in $\pi \phi^0$ and $\omega \pi^0$ are $< 7\%$ and $\sim
37\%$, respectively, and $\phi$ and $\omega$ production have different
energy dependences in $n \bar{p}$ annihilations. Also, it has been
observed that the initial states in $p \bar{p} \to \phi \phi$ are
dominated by $J^{PC} = 2^{++}$, consistent with S-wave annihilations in a
spin-triplet state. Additionally, spin-singlet initial states are strongly
suppressed in $p \bar{p} \to \Lambda \bar{\Lambda}$: the singlet fraction
$F_s = (0.1 \pm 7.3) \times 10^{-3}$. The polarized-strangeness model is
also consistent with the available data on the Pontecorvo reaction
$\bar{p} d \to \phi n$ and on selection rules in $p \bar{p} \to K^* K^*$.
Other successful predictions include `OZI violation' in nucleon-nucleon
scattering, where $p \bar{p} \to p \bar{p}\phi$ is about 14 times more
copious than $p \bar{p} \to p \bar{p} \omega$ near threshold and the
$\phi$ and $\omega$ angular distributions are different, the `violation'
of the na\"{i}ve OZI rule by a factor $\sim 20$ in $ pd \to ^3H_e
\phi(\omega)$, and the negative longitudinal polarization of $\Lambda$
baryons measured in deep-inelastic neutrino scattering~\cite{NOMAD}, 
discussed below.

However, there are also some serious problems for the
polarized-strangeness model. For example, the strong OZI `violation' in
$\bar{p} p \to \gamma \phi$ takes place from a $^1S_0$ initial state, and
the spin transfer $D_{nn}$ in $p \bar{p} \to \Lambda \bar{\Lambda}$ is
small, whereas $K_{nn} > 0$, indicating that the spin of the proton is
transferred to the $\bar{\Lambda}$, not to the
$\Lambda$~\cite{LEARLambda}. Moreover, CLAS data on the reaction $\vec{e}
p \to e' K^{+} \Lambda$ indicate that the spins of the $s$ and $\bar{s}$
are anti-aligned~\cite{CLASLambda}. Also serious is the problem that 
${\bar p} p
\to \pi^0 \phi$ is not possible from a $^3S_1$ initial state without 
either
flipping the spin of the $s$-quark or positive polarization of the strange
quarks in the proton~\cite{Rekalo}.

Many of these problems would be resolved if there are two components of
polarized strangeness, one with $S_z = -1$ and one with $S_z =
0$~\cite{both}. This would permit $p \bar{p} \to \gamma \phi$ and $p
\bar{p} \to \pi^0 \phi$ via rearrangement diagrams, and the CLAS data that
require the spins of the $s$ and $\bar{s}$ to be anti-aligned could be
accommodated by shake-out of the $S_z = 0$ component. However, even this
model does not fit all the data, as seen in the Table. One 
promising possibility is to assume the dominance of a 
spin-singlet $us$ diquark configuration, as indicated in the last column 
of the Table~\cite{both}. Understanding 
the strange polarization of the proton is still a work in progress.

\begin{table*}
\hspace*{3.5cm}
\begin{tabular}{|l|l|l|l|}
\hline\noalign{\smallskip}
& $0^{++}: S_z = -1$ & $0^{++}: S_z =-1,0$ & $0^{-+}: (us), (s\bar{s})$ \\
\noalign{\smallskip}\hline\noalign{\smallskip}
$\phi \pi / \omega \pi$: {\rm large} & + & - & +\\
{\rm from} $^3S_1$ & & & \\ \hline
$\phi \pi$: spin state & - & + & + \\ \hline
$\phi \gamma / \omega \gamma$: {\rm large} & - & + & glueball \\
{\rm from} $^1S_1$ & & &  \\ \hline
$\phi \eta / \omega \eta$: {\rm small} & {\rm small} \, $Q^2$ & {\rm 
small}
\, $Q^2$ & {\rm small} \, $Q^2$ \\
{\rm from} $^3S_1$ & & & \\ \hline
$\phi \rho / \omega \rho$: {\rm small} & + & - & +\\
{\rm from} $^1S_1$ & & &  \\ \hline
$f'_2 / f_2$: {\rm large} & + & - & + \\
{\rm from $p$-wave} &&& \\ \hline
$\phi n / \omega n$: {\rm large} & {\rm large} \, $Q^2$ & {\rm large} \,
$Q^2$ & {\rm large} \, $Q^2$ \\ \hline
$P ( \Lambda ) < 0$ {\rm in DIS} & + & + & + \\ \hline
$ep \to \Lambda Ke: P(\Lambda)$ & - & + & + \\ \hline
$\bar{p} p \to \Lambda \bar{\Lambda}: D_{nn}$ & - & + &+/- \\
\hline
$\bar{p} p \to \Lambda \bar{\Lambda}: K_{nn}$ & - & - & + \\
\hline
$pp \to pp \phi$: {\rm large} & + & - & + \\
{\rm from} $^3S_1$ & & & \\
\noalign{\smallskip}\hline
\end{tabular}
\caption{Score card for various models of polarized strangeness in the 
nucleon wave function.}
\end{table*}

\section{Probing Strangeness via $\Lambda$ Polarization}

Since the polarization of the $\Lambda$ is measurable in its decays, and
since the $\Lambda$ polarization is inherited, at least in the na\"{i}ve
quark model, from its constituent $s$ quark, $\Lambda$ polarization is
potentially a powerful way of probing polarized strangeness. Particularly
interesting from this point of view is the measurement of $\Lambda$
polarization in leptoproduction, where two options are available:
measurements in the fragmentation region of the struck quark or in that of
the target. The struck quark has net polarization, but is usually a $u$,
so there is no interesting spin transfer to the $\Lambda$ baryon. However,
in the target fragmentation region the `wounded nucleon' left behind by
the polarized struck quark is itself polarized in general. A priori, it is
a diquark system with the possibility of a polarized $s \bar{s}$ `sea'
attached to it. Memory of this polarization may be carried by the $s$ and
$\bar{s}$ in the wounded nucleon wave function and transferred to
$\Lambda$ and $\bar{\Lambda}$ baryons produced in the target fragmentation
region~\cite{EKS}.

\begin{figure}
\resizebox{0.45\textwidth}{!}{%
\includegraphics{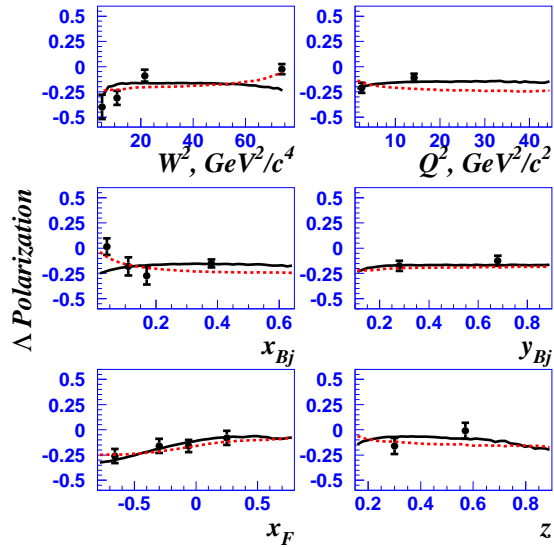}
}
\caption{Predictions for longitudinal $\Lambda$ polarization in 
deep-inelastic scattering~\protect\cite{EKN}, compared with data from 
NOMAD~\protect\cite{NOMAD}.
}
\label{fig:Naumov}
\end{figure}

We have modelled this idea using the Lund string fragmentation model
incorporated in LEPTO~6.5.1 and JETSET~7.4, and have considered various
combinations of two extreme cases in which the $\Lambda$ baryon is
produced by fragmentation of either the struck quark or the remnant
diquark~\cite{EKN}. We then fix free parameters of the model by demanding
consistency with data from NOMAD in deep-inelastic $\nu$
scattering~\cite{NOMAD}. In addition to providing a good fit to NOMAD
data, as seen in Fig.~\ref{fig:Naumov}, this procedure can be used 
directly to make 
predictions for
electroproduction data from HERMES, and agrees very well. We have then
gone on to make predictions for the COMPASS muon scattering experiment.
COMPASS was originally conceived to measure the polarization of the gluons
in the proton, but it may also be able to cast light on the polarization
of the strange quarks!

\section{Summary}

As we have seen in this review, there are many pieces of experimental
evidence for a significant amount of hidden strangeness in the proton wave
function, notably the $\pi$-nucleon $\Sigma$ term, polarized
deep-inelastic scattering and large deviations from the na\"{i}ve OZI
rule. These observations may cast light on complementary models of nucleon
structure, namely the `na\"{i}ve' quark model and chiral soliton models.
The latter were recently boosted by reports of exotic baryons, whose 
existence was
predicted years ago in the soliton model. Their spectroscopy depends, in
particular, on the magnitude of the $\pi$-nucleon $\Sigma$ term, and the
tentative indication from the difference between the masses of the
$\theta^{+}$ and $\Xi^{--}$-baryons is that this should be large, in
agreement with the latest direct determinations of this quantity. The
situation with phenomenological models of OZI `evasion' due to polarized
$s \bar{s}$ pairs in the nucleon wave function is unclear: the data from
LEAR and other low-energy experiments suggests that there must be many $s
\bar{s}$ pairs, but their polarization states remain obscure. One thing
is, however, clear: we may expect many more twists in the strange story of
the nucleon!

\section*{Acknowledgements}
It is a pleasure to thank my collaborators on the subjects discussed here, 
namely S.~J.~Brodsky, E.~Gabathuler, M.~Karliner, D.~E.~Kharzeev, 
A.~Kotzinian, D.~V.~Naumov and M.~G.~Sapozhnikov.

\end{document}